\documentclass[]{jetpl}
\twocolumn
\usepackage{epsf}

\lat

\renewcommand{\vec}{\mathbf}
\renewcommand{\d}{\mathrm d}

\title{Mesoscopic wave turbulence.}

\rtitle{Mesoscopic wave turbulence.}

\sodtitle{Mesoscopic wave turbulence.}

\author{V.\,E.\,Zakharov$^{+*\ddag}$, 
A.\,O.\,Korotkevich$^{+}$\/\thanks{e-mail: kao@landau.ac.ru}, A.\,N.\,Pushkarev$^{+\ddag}$,
A.\,I.\,Dyachenko$^{+}$}

\rauthor{V.\,E.\,Zakharov, A.\,O.\,Korotkevich, A.\,Pushkarev, A.\,I.\,Dyachenko}

\sodauthor{Zakharov, Korotkevich, Pushkarev, Dyachenko}

\address{$^+$L.D. Landau Institute for Theoretical Physics RAS,
2 Kosygin Str., Moscow, 119334 Russian Federation\\~\\
$^*$Department of Mathematics, University of Arizona, Tucson, AZ 85721 USA\\~\\
$^\ddag$Waves and Solitons LLC, 738 W. Sereno Dr., Gilbert, AZ 85233 USA
}
\dates{16 August 2005}{*}
\abstract{We report results of sumulation of wave turbulence. Both inverse and direct cascades are observed.
The definition of "mesoscopic turbulence" is given. This is a regime when the number of modes in a system
involved in turbulence is high enough to qualitatively simulate most of the processes but significantly smaller then the
threshold, which gives us quantitative agreement with the statistical description, such as kinetic equation. Such a regime
takes place in numerical simulation, in essentially finite systems, etc.}
\begin{document}
\maketitle

The theory of wave turbulence is developed for infinitely large system. In weakly nonlinear dispersive
media, the turbulence is described by a kinetic equation for squared wave amplitudes (weak turbulence).
However, all real systems are finite. Computer simulation of wave turbulence can also be perfomed only in finite
system (typically in a box with periodic boundary conditions). It is important to know how strong discreteness
of a system impacts the physical picture of wave turbulence.

Let a turbulence be realized in a Q-dimensional cube with side L. Then, wave vectors form a cubic lattice
with the lattice constant $\Delta k = 2\pi/L$. Suppose that four-wave resonant conditions are dominating.
Exact resonances satisfy the equations
\begin{equation}
\label{Diophant1}
\vec k + \vec k_1 - \vec k_2 - \vec k_3 = 0,
\end{equation}
\begin{equation}
\label{Diophant2}
\Delta = \omega (k) + \omega (k_1) - \omega (k_2) - \omega (k_3) = 0.
\end{equation}
In infinite medium, Eqs. (\ref{Diophant1}) and (\ref{Diophant2}) define hypersurface of dimension $3Q-1$ in $4Q$-dimensional space
$\vec k, \vec k_1, \vec k_2, \vec k_3$. In a finite system, (\ref{Diophant1}) and (\ref{Diophant2}) are Diophantine equations which
might have or have no exact solutions. The Diophantine equation for four-wave resonant
processes are not studied yet. For three-wave resonant processes, they are studied for
Rossby waves on the $\beta$-plane~\cite{Kartashova-3}.

However, not only exact resonances are important. Individual harmonics in the wave ensemble fluctuate
with inverse time $\Gamma_{\vec k}$, dependent on their wavenumbers. Suppose that all $\Gamma_{\vec k_i}$
for waves, composing a resonant quartet, are of the same order of magnitude $\Gamma_{\vec k_i} \sim \Gamma$.
Then resonant equation (\ref{Diophant2}) has to be satisfied up to accuracy $\Delta \sim \Gamma$, and the
resonant surface is blurred into the layer of thickness $\delta k / k \simeq \Gamma_k / \omega_k$. This thickness
should be compared with the lattice constant $\Delta k$. Three different cases are possible
\begin{enumerate}
\item $\delta k \gg \Delta k$.
In this case the resonant layer is thick enough to hold many approximate resonant quartets on a unit of resonant
surface square. These resonances are dense, and the theory is close to the classical weak turbulent theory
in infinite media.
The weak turbulent theory offers recipes for calculation of $\Gamma_k$. The weak-turbulent $\Gamma_k$
are the smallest among all given by theoretical models. To be sure that the case is realized, one has to use
weak-turbulent formulae for $\Gamma_k$.
\item $\delta k < \Delta k$.
This is the opposite case. Resonances are rarefied, and the system consists of a discrete set of weakly
interacting oscillators. A typical regime in this situation is the "frozen
turbulence"~\cite{Pushkarev-1999}, \cite{Pushkarev-1999a}, \cite{Pushkarev-2001},
which is actually a system of KAM tori, accomplished with a weak Arnold's diffusion.
\item The intermediate case $\delta k \simeq \Delta k$ can be called "mesoscopic turbulence".
The density of approximate resonances is high enough to provide the energy transport along the spectrum,
but low enough to guarantee "equal rights" for all harmonics, which is a necessary condition for applicability of
the weak turbulent theory.
\end{enumerate}

In this article we report results of our numerical experiments on modeling of turbulence of gravity waves on
the surface of deep ideal incompressible fluid.
The motivation for this work was numerical justification of Hasselmann kinetic equation.
The result is discovery of the mesoscopic turbulence.
The fluid motion is potential and described by shape of surface
$\eta(\vec r, t)$ and velocity potential $\psi (\vec r, t)$, evaluated on the surface. These variables satisfy the
canonical equations~\cite{Zakharov-1968}
\begin{equation}
\label{Hamiltonian_equations}
\frac{\partial \eta}{\partial t} = \frac{\delta H}{\delta \psi}, \;\;\;\;
\frac{\partial \psi}{\partial t} = - \frac{\delta H}{\delta \eta},
\end{equation}
Hamiltonian $H$ is presented by the first three terms in expansion on powers of nonlinearity $\nabla \eta$
\begin{equation}
\label{Hamiltonian}
\begin{array}{l}
\displaystyle
H = H_0 + H_1 + H_2 + ...,\\
\displaystyle
H_0 = \frac{1}{2}\int\left( g \eta^2 + \psi \hat k  \psi \right) \d x \d y,\\
\displaystyle
H_1 =  \frac{1}{2}\int\eta\left[ |\nabla \psi|^2 - (\hat k \psi)^2 \right] \d x \d y,\\
\displaystyle
H_2 = \frac{1}{2}\int\eta (\hat k \psi) \left[ \hat k (\eta (\hat k \psi)) + \eta\nabla^2\psi \right] \d x \d y.
\end{array}
\end{equation}
Thereafter, we put gravity acceleration equal to $g=1$. Here, $\hat k$ is a linear integral operator 
$\left(\hat k =\sqrt{-\nabla^2}\right)$, such that, in $k$-space, it corresponds to
multiplication of Fourier harmonics ($\psi_{\vec k} = \frac{1}{2\pi} \int \psi_{\vec r} e^{i {\vec k} {\vec r}} \d x \d y$)
by $\sqrt{k_{x}^2 + k_{y}^2}$. For gravity waves, this
reduced Hamiltonian describes four-wave interaction. Then, dynamical equations (\ref{Hamiltonian_equations}) acquire the form
\begin{equation}
\label{eta_psi_system}
\begin{array}{lcl}
\displaystyle
\dot \eta &=& \hat k  \psi - (\nabla (\eta \nabla \psi)) - \hat k  [\eta \hat k  \psi] +\\
\displaystyle
		&&+ \hat k (\eta \hat k  [\eta \hat k  \psi]) + \frac{1}{2} \nabla^2 [\eta^2 \hat k \psi] + 
		\frac{1}{2} \hat k [\eta^2 \nabla^2\psi],\\
\displaystyle
\dot \psi &=& - g\eta - \frac{1}{2}\left[ (\nabla \psi)^2 - (\hat k \psi)^2 \right] - \\
\displaystyle
		&& - [\hat k  \psi] \hat k  [\eta \hat k  \psi] - [\eta \hat k  \psi]\nabla^2\psi.
\end{array}
\end{equation}
Let us introduce the canonical variables $a_{\vec k}$ as shown below
\begin{equation}
a_{\vec k} = \sqrt \frac{\omega_k}{2k} \eta_{\vec k} + i \sqrt \frac{k}{2\omega_k} \psi_{\vec k},
\end{equation}
where $\omega_k = \sqrt {gk}$. In these so called normal variables equations (\ref{Hamiltonian_equations}) take
the form
\begin{equation}
\frac{\partial a_{\vec k}}{\partial t} = -i\frac{\delta H}{\delta a_{\vec k}^{*}}.
\end{equation}
The physical meaning of these variables is quite clear: $|a_{\vec k}|^2$ is an action spectral density,
or $|a_{\vec k}|^2\Delta k ^2$ is a number of particles with the particular wave number ${\vec k}$.

We solved equations (\ref{eta_psi_system}) numerically in a box $2\pi \times 2\pi$
using a spectral code on rectangular grid with double
periodic boundary conditions. The implicit energy-preserving scheme, similar to used in~\cite{Capillary-2003},
\cite{Gravity-2003}, \cite{Gravity-2004}, was implemented. We studied evolution of freely propagating
waves (swell) in the absence of wind in the spirit of paper~\cite{Onorato-2002}. Different grids ($512\times512$,
$256\times1024$, $256\times2048$) with different initial data were tried. In all the cases, we observed mesoscopic
wave turbulence. The most spectacular results are achieved on the grid $256\times2048$.

As initial conditions, we used a Gauss-shaped distribution on a long axis of the wavenumbers plane
\begin{equation}
\begin{array}{l}
\displaystyle
\left\{
\begin{array}{l}
\displaystyle
|a_{\vec k}| = A_i \exp \left(- \frac{1}{2}\frac{\left|\vec k - \vec k_0\right|^2}{D_i^2}\right), 
\left|\vec k - \vec k_0\right| \le 2D_i,\\
\displaystyle
|a_{\vec k}| = 10^{-12}, \left|\vec k - \vec k_0\right| > 2D_i,
\end{array}
\right.\\
\displaystyle
A_i = 5\times10^{-6}, D_i = 30, \vec k_0 = (0; 150).
\end{array}
\end{equation}
The initial phases of all the harmonics were random. The average steepness $\mu = <|\nabla \eta|> \simeq 0.115$. To
stabilize the computations in the high-frequency region~\cite{Lushnikov-Zakharov-2005}, we introduced artificial damping, mimicking viscosity
at small scales, and an artificial smoothing term to the equation for the surface evolution
\begin{equation}
\begin{array}{l}
\displaystyle
\frac{\partial \psi_{\vec k}}{\partial t} \to \frac{\partial \psi_{\vec k}}{\partial t} + \gamma_k\psi_{\vec k},\\
\displaystyle
\frac{\partial \eta_{\vec k}}{\partial t} \to \frac{\partial \eta_{\vec k}}{\partial t} + \gamma_k\eta_{\vec k},\\
\displaystyle
\gamma_k = \left\{
\begin{array}{l}
\displaystyle
0, k < k_d,\\
\displaystyle
- \gamma (k - k_d)^2, k \ge k_d,\\
\end{array}
\right.\\
\displaystyle
k_d = 512, \gamma = 2 \times 10^{4}, \tau = 3.1\times10^{-4}.
\end{array}
\end{equation}
With the  time step $\tau$, this calculations took about two months on AMD Athlon 64 3500+ computer. During this time, we
reached 1500 periods of the wave in the initial spectral maximum.

The process of waves evolution can be separated in two steps. On the first stage (about fifty initial wave periods),
we observe fast loss of energy and wave action. This effect can be explained by formation of "slave" harmonics
taking their part of motion constants. Initially smooth spectrum becomes very rough.
The spectral maximum demonstrates fast
downshift.

\begin{figure}[htb!]
\centering
\includegraphics[width=8.0cm]{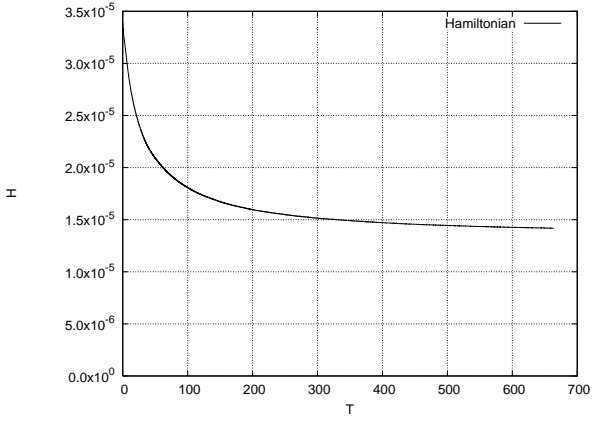}
\caption[]{\label{Hamiltonian_vs_time}Fig.\ref{Hamiltonian_vs_time}. Total energy of the system.}
\end{figure}

\begin{figure}[htb!]
\centering
\includegraphics[width=8.0cm]{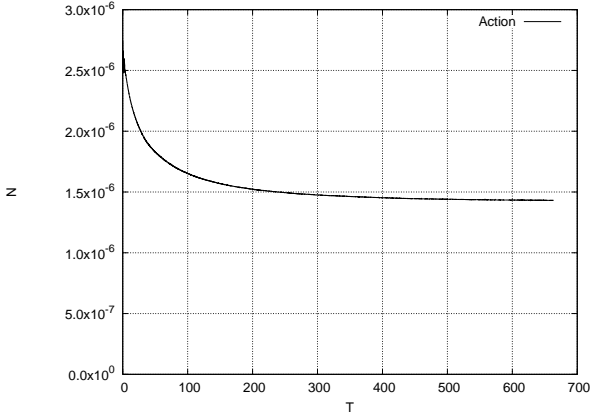}
\caption[]{\label{action_vs_time}Fig.\ref{action_vs_time}. Total action of the system.}
\end{figure}

\begin{figure}[htb!]
\centering
\includegraphics[width=8.0cm]{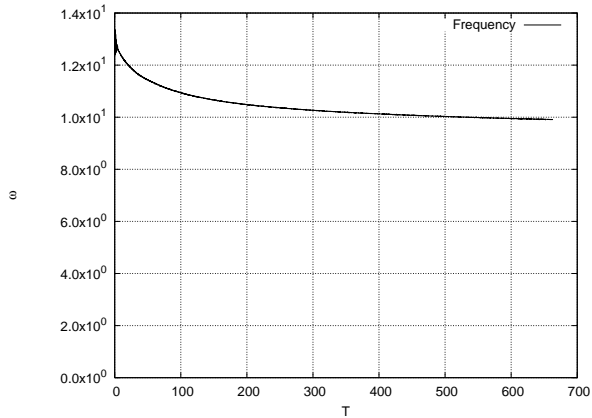}
\caption[]{\label{frequency_vs_time}Fig.\ref{frequency_vs_time}. Frequency of the spectral maximum.}
\end{figure}

\begin{figure}[htb!]
\centering
\includegraphics[width=8cm]{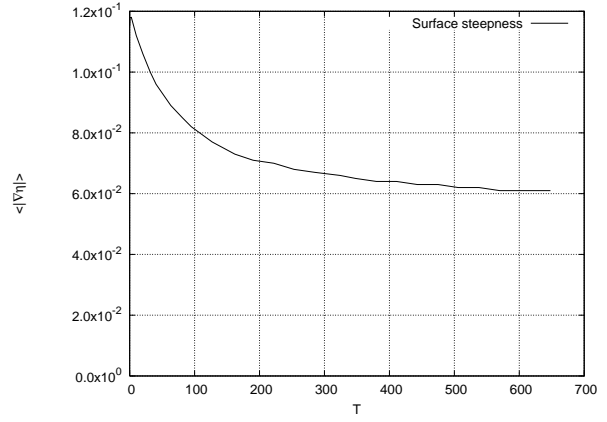}
\caption[]{\label{steepness_vs_time}Fig.\ref{steepness_vs_time}. Mean steepness of fluid surface.}
\end{figure}

\begin{figure}[htb!]
\centering
\includegraphics[width=8cm]{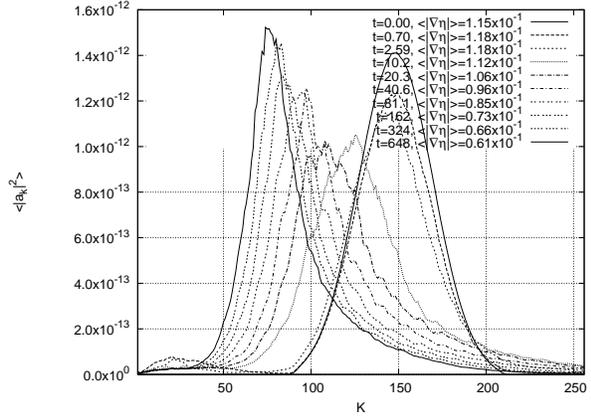}
\caption[]{\label{a_k_anglemov} Fig.\ref{a_k_anglemov}. Averaged with angle spectra. Downshift of spectral  maximum is clearly observable.}
\end{figure}

\begin{figure}[htb!]
\centering
\includegraphics[width=8cm]{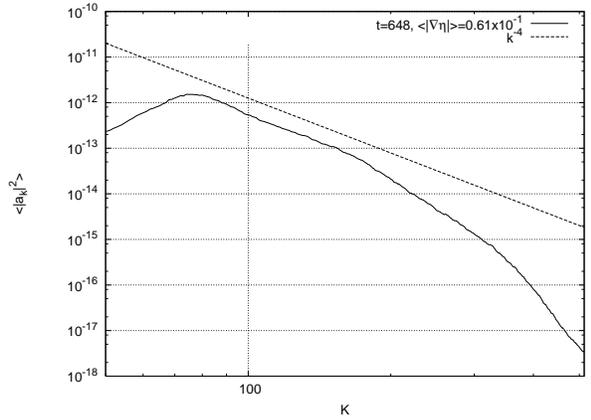}
\caption[]{\label{Kolmogorov} Fig.\ref{Kolmogorov}.Tails of angle-averaged spectrum in
double logarithmic scale. $T=648=1263T_0$. Power-like tail and front slope are close to predicted by the weak
turbulent theory.}
\end{figure}

In the second stage, the downshift continues but all processes slow down. Plots of energy, wave action,
mean frequency, and mean steepness are presented in Figs. \ref{Hamiltonian_vs_time}-\ref{steepness_vs_time}.
One can see clear tendency to downshift of the spectral maximum corresponding to inverse cascade; however, this
process is more slow than predicted by the weak turbulence theory. Self-similar downshift in this theory
gives~\cite{Zakharov-PhD}, \cite{Zakharov-1982}
$$
\omega \sim t^{-1/11}.
$$
In our experiments
$$
\omega \sim t^{-\alpha},
$$
where $\alpha$ decreases with time from $1/16$ to $1/20$.
Evolution of angle averaged spectra $N_k = \int \limits_{0}^{2\pi} |a_{\vec k}|^2 k \d k \d \vartheta$ is presented
on Fig. \ref{a_k_anglemov}.
Their tails (Fig. \ref{Kolmogorov}) are Zakharov-Filonenko weak-turbulent
Kolmogorov spectra \cite{Zakharov-JAMTP67} corresponding to direct cascade
\begin{equation}
|a_k|^2 \sim \frac{1}{k^4}.
\end{equation}
This result is robust; it was observed in similar calculations \cite{Onorato-2002}, 
\cite{Gravity-2003}, \cite{Gravity-2004}.

\begin{figure}[htb!]
\centering
\includegraphics[width=8.5cm]{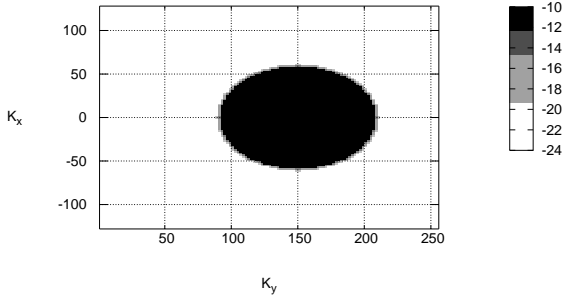}
\caption[]{\label{a_k_c_initial} Fig.\ref{a_k_c_initial}. Level lines of logarithm of initial spectra distribution. $T=0$.}
\end{figure}

\begin{figure}[htb!]
\centering
\includegraphics[width=8.5cm]{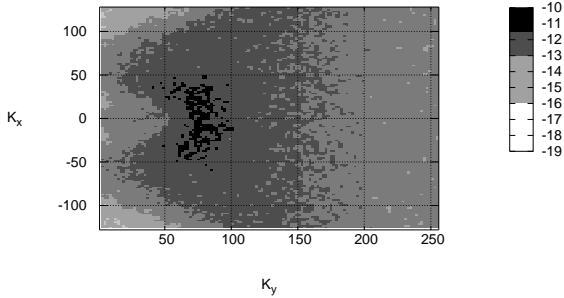}
\caption[]{\label{a_k_c_final} Fig.\ref{a_k_c_final}. Level lines of logarithm of spectra distribution at $T=648=1263T_0$.}
\end{figure}

Two dimensional spectra in the initial and in the last moments of calculations are presented in Fig.
\ref{a_k_c_initial}, \ref{a_k_c_final}. One can see formation of small intensity "jets" posed on the Phillips
resonant curve~\cite{Phillips-1981}
\begin{equation}
\label{Phillips_curve}
2\omega(\vec k_0) = \omega(\vec k_0 + \vec k) + \omega(\vec k_0 - \vec k).
\end{equation}

\begin{figure}[htb!]
\centering
\includegraphics[width=8cm]{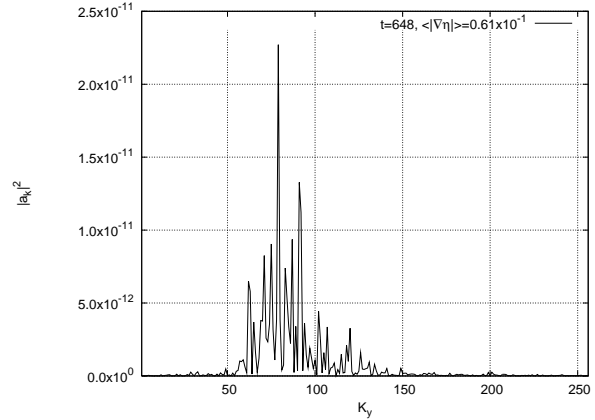}
\caption[]{\label{Spectrum_cut} Fig.\ref{Spectrum_cut}. Slice of spectrum
on axis $(0; k_y)$ at $T=648=1263T_0$.}
\end{figure}

\begin{figure}[htb!]
\centering
\includegraphics[width=8cm]{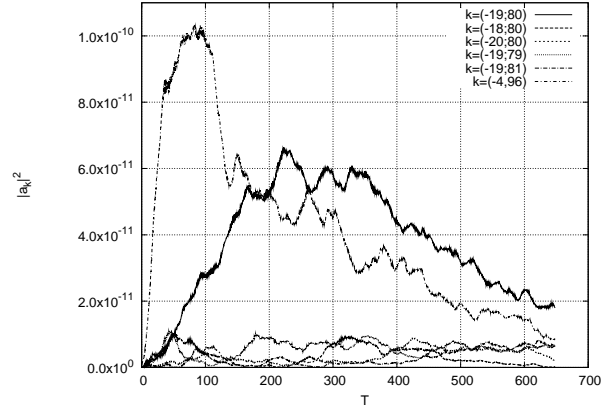}
\caption[]{\label{multiple_harmonics} Fig.\ref{multiple_harmonics}. Evolution of some
cluster of harmonics and a distant large harmonic.}
\end{figure}

The spectra are very rough and sharp. The slice of spectra along the line $(0; k_y)$ in the end of the computations is
presented on Fig. \ref{Spectrum_cut}. Evolution of squared wave amplitudes for a cluster of neighbouring
harmonics is presented in Fig. \ref{multiple_harmonics}.

Results presented in Fig. \ref{multiple_harmonics} show that what we modeled is mesoscopic turbulence.
Indeed, characteristic time of amplitude evolution on a figure is a hundred or more their periods; thus
$\Gamma/\omega_k$ is comparable with $\Delta k /k$. On the same figure we can see the most remarkable
features of such turbulence.

\begin{figure}[htb!]
\centering
\includegraphics[width=8cm]{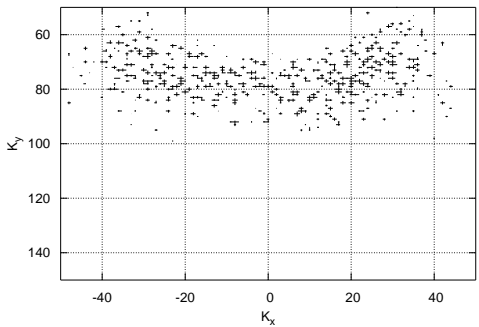}
\caption[]{\label{Champions} Fig.\ref{Champions}. Harmonics with square modulus
exceeding the level $10^{-11}$ at $T=648=1263T_0$.}
\end{figure}

The weak turbulence in the first approximation obeys the Gaussian statistics. The neighbouring harmonics are
uncorrelated and statistically independent ($\left<a_k a_{k+1}^*\right> = 0$). However, their averaged characteristics are
close to each other. This is a "democratic society". On the contrary mesoscopic turbulence is an "oligarchicÓ
society". The Phillips curve (\ref{Phillips_curve}) has a genus 2. After Faltings' proof~\cite{Faltings-1983}
of Mordell's hypothesis~\cite{Mordell-1922} we know that the number of solutions of the Diophantine equation
\begin{equation}
\begin{array}{c}
\displaystyle
\Delta = 2 (n^2 + m^2)^{1/4} - [(n + x)^2 + (m + y)^2]^{1/4} -\\
\displaystyle
- [(n - x)^2 + (m - y)^2]^{1/4} = 0
\end{array}
\end{equation}
is at most finite and most probably, except for a few trivial solutions, equals to zero.
The same statement is very plausible for more general resonances. Approximate integer solutions
in the case
$$
|\Delta| < \epsilon
$$
do exist, but their number fast tends to zero at $\epsilon \to 0$. Classification of these solutions is a hard
problem of the number theory. These solutions compose the "elite society" of the harmonics, which play
the most active role in the mesoscopic turbulence. Almost all the inverse cascade of wave action is realized within
members of this "privileged club". The distribution of the harmonics exceeding the reference level $|a_k|^2 = 10^{-11}$
at the moment $t=1200T_0$ is presented in Fig. \ref{Champions}. The number of such harmonics is not more
than 600, while the total number of harmonics involved into the turbulence is of the order of $10^{4}$.

Note that a situation with direct cascade is different. As far as the coupling coefficient for gravity waves growth
as fast as $k^3$ with the wave number, for short waves $\Gamma_k/\omega_k$ easily exceeds $\Delta k /k$,
and the conditions of applicability of the weak turbulent theory for short waves are satisfied.

Note also that the mesoscopic turbulence is not a numerical artefact. Simple estimations show that, for
gravity waves, it is realized in some conditions in basins of a moderate size, like small lakes as well as in
experimental wave tanks. It is also common for long internal waves in the ocean and for inertial gravity waves
in atmosphere, for plasma waves in tokamaks, etc.

This work was
supported by RFBR grant 03-01-00289, the Programme
``Nonlinear dynamics and solitons'' of the RAS Presidium and ``Leading Scientific
Schools of Russia" grant,
also by ONR grant N00014-03-1-0648, 
US Army Corps of Engineers, RDT\&E Programm W912HZ-04-P-0172,
Grant DACA 42-00-C0044.
We use this opportunity to gratefully acknowledge the support of
these foundations.

Also, the authors want to thank creators of the opensource fast Fourier transform library
FFTW~\cite{FFTW} for this fast, portable and completely free piece of software.

\end{document}